\documentclass{camera}

\begin{document}

%
\title{QUANTUM POINCAR\'E RECURRENCES IN MICROWAVE IONIZATION
OF RYDBERG ATOMS}

%
\author{Giuliano Benenti \And Giulio Casati} 

%
\organization{Universit\`a degli Studi dell'Insubria and 
Istituto Nazionale per la Fisica della Materia, via Lucini 3, 
22100 Como, Italy}  

\maketitle

%

\input{epsf}

\begin{abstract}
We study the time dependence of the ionization probability of 
Rydberg atoms driven by a microwave field. 
The quantum survival probability follows the classical one 
up to the Heisenberg time and then decays inversely proportional 
to time, due to tunneling and localization effects.   
We provide parameter values which should allow one to observe such  
decay in laboratory experiments. 
Relations to the $1/f$ noise are also discussed.  
\end{abstract}

The pioneering experiment of Bayfield and Koch performed in 1974 \cite{BK}
attracted a great  interest to ionization of highly excited
hydrogen and Rydberg atoms in a microwave field.  
The main reason of this
interest is due to the fact that such ionization
requires the absorption of a large number of  photons (about $20-70$)
and can be explained only as a result of the appearence of dynamical chaos and 
diffusive energy excitation in the corresponding
classical system. 
Such classical diffusive ionization
requires many microwave periods 
and quantum interference effects can suppress this diffusion
leading to quantum localization of chaos \cite{IEEE}.

More recent experiments \cite{Delande} with alkali Rydberg atoms showed 
an algebraic time dependence of the survival probability:    
$P(t)\propto t^{-\alpha}$, with $\alpha\approx 0.5$. 
This result cannot be explained  
within the picture of diffusive ionization in the domain of 
classical chaos. 
The origin of the slow algebraic decay was attributed to the 
underlying structure of classical mixed phase space composed by 
integrable islands surrounded by chaotic components:  
Chaotic trajectories can be trapped in the vicinity of the hierarchy 
of regular islands \cite{CS99} and this slows down the ionization process.  

Recent studies of quantum Poincar\'e recurrences for 
the standard map model in the semiclassical regime \cite{QPR}
showed that quantum $P(t)$ follows the classical decay up to the 
Heisenberg time $t_H$, which is determined by inverse level spacings. 
For $t>t_H$, the quantum survival probability 
starts to decay inversely proportional to time ($\alpha=1$) and becomes 
much larger than the classical one. The power $\alpha=1$ is due to 
the exponentially broad distribution of time scales to escape 
from some region of phase 
space \cite{QPR}. These exponentially broad escape times are originated 
by tunneling from classically integrable regions or by the exponential 
quantum localization. 

Microwave ionization of Rydberg atoms constitutes a unique 
possibility to experimentally observe quantum Poincar\'e recurrences \cite{hqpr}. 
Indeed the highly excited states of hydrogen atom in a microwave field 
can be approximately described by the Kepler map which is very similar 
to the standard map \cite{IEEE} and therefore one would expect the 
same behavior for the time dependence of the survival probability. 

In order to investigate the probability decay for the hydrogen atom in a 
microwave field, we choose the initial state with principal quantum number 
$n_0$ and numerically study the survival probability $P(t)$ in a 
linearly polarized monochromatic electric field. 
The Hamiltonian reads 
\begin{equation} 
H=\frac{p^2}{2}-\frac{1}{r}+\epsilon z \sin (\omega t),  
\end{equation} 
wher $\epsilon$ and $\omega$ are the 
strength and frequency of the microwave field, measured in atomic 
units. 
The quantum evolution is numerically simulated  
by the one-dimensional ($1d$) model of a hydrogen atom and by the 
$3d$ model for atoms initially prepared in states extended along  
the field direction and with magnetic quantum number $m=0$. 
In the $1d$ model the motion is assumed to take place along the 
field direction ($z$-axis, with $z\geq 0$). 

In order to compare classical and quantum dynamics it is convenient 
to use the scaled field strength $\epsilon_0=\epsilon n_0^4$ and frequency 
$\omega_0=\omega n_0^3$, which completely determine the classical 
dynamics. The classical limit corresponds to $\hbar_{\rm eff}=   
\hbar/n_0\to 0$, at constant $\epsilon_0$, $\omega_0$. 
In classical mechanics diffusive ionization  
takes place for fields above the chaos border: 
$\epsilon_0>\epsilon_c\approx 1/(49\omega_0^{1/3})$ 
\cite{IEEE}. Quantum interference effects can suppress this 
diffusion leading to quantum localization of chaos \cite{IEEE}. Such 
dynamical localization leads to a quantum probability distribution 
$f_N$ exponentially localized in the number of absorbed photons 
$N_\phi=(E-E_0)/\omega$ ($E$ electron energy, $E_0=-1/2n_0^2$): 
$f_N\propto\exp(-2|N_\phi|/\ell_\phi)$, with localization length  
$\ell_\phi=3.3\epsilon_0^2 \omega_0^{-10/3} n_0^2$ \cite{IEEE}. 

We introduce an absorption border for levels with $n\geq n_c$; 
like this the number of photons required to ionize the atom 
is $N_c=(n_0/2\omega_0)(1-n_0^2/n_c^2)$.  
Such  border occurs in real laboratory experiments, for example as 
a consequence of unavoidable stray electric fields experienced 
by the Rydberg atoms during their interaction with the microwave.
The absorption border $n_c$ can be varied in a controlled way via a 
static electric field $\epsilon_s$, the static field ionization 
border being $\epsilon_s n_c^4\approx 0.13$. 

\begin{figure} 
\centerline{\epsfxsize=10cm\epsffile{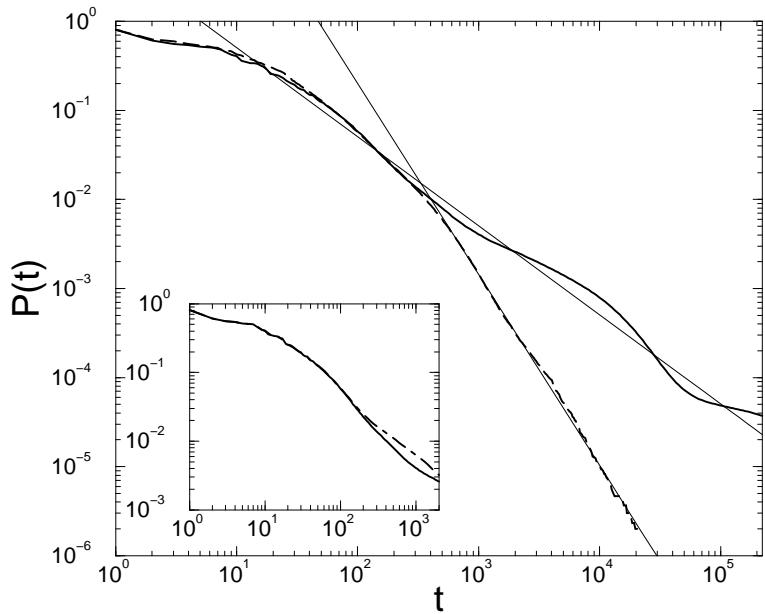}}
\caption{\small Survival probability $P(t)$ as a function of 
the interaction time $t$ (in units of microwave periods) 
for $\epsilon_0=0.1$, $\omega_0=2.6$, $n_0=60$, $n_c=64$: 
quantum (solid line) and classical (dashed 
line, ensemble of $3\times 10^6$ trajectories) 
$1d$ hydrogen model. The straight lines have slopes 
$1$ and $2.15$, the latter coming from a fit of the classical 
decay for $5\times  10^2<t<2\times 10^4$. Inset: quantum  
survival probability for the $1d$ model (solid line) and the 
$3d$ hydrogen atom (dot-dashed line).}  
\label{fig1} 
\end{figure}

In Fig. \ref{fig1} we show a realistic case ($n_0=60$, $\epsilon_0=0.1$,
$\omega_0=2.6$) in which, initially, 
classical and quantum probabilities decay in a very similar way and 
only after approximately $5\times 10^2$ microwave periods the quantum 
survival probability starts to decay more slowly ($P(t) \propto 1/t$) than 
the classical one which decays as $1/t^{\alpha}$, with $\alpha\approx 2.15$. 
The comparison of quantum simulations for the $1d$ hydrogen 
atom model and  the $3d$ dynamics is shown in the inset of Fig. \ref{fig1}. 
It demonstrates that both dynamics give very close results, confirming  
that the essential physics is captured by the $1d$ model.   
Actually, due to Coulomb degeneracy, the slow motion in the orbital 
momentum $l$ acts as an adiabatic perturbation on the $n$ motion 
and as a result the excitation in $n$ is well described by the $1d$ 
model \cite{IEEE}. 
We put the absorption border  near  the initial state ($n_c=64$) in 
order to have $\rho_c=\ell_\phi/ N_c \approx 3.5 > 1$.
In this  way the probability can 
go out very easily and the $1/t$ probability decay is observed after a 
short transient time of the order of $20$ microwave periods. 
On the contrary, when $\rho_c <1$, strong fluctuations around the 
$1/t$ decay take place \cite{hqpr}. This is analogous to the huge 
(log-normally distributed) conductance fluctuations 
in a disorder solid with localization length smaller 
than the sample size.  

\begin{figure} 
\vspace{-0.5cm}
\centerline{\epsfxsize=5.0cm\epsffile{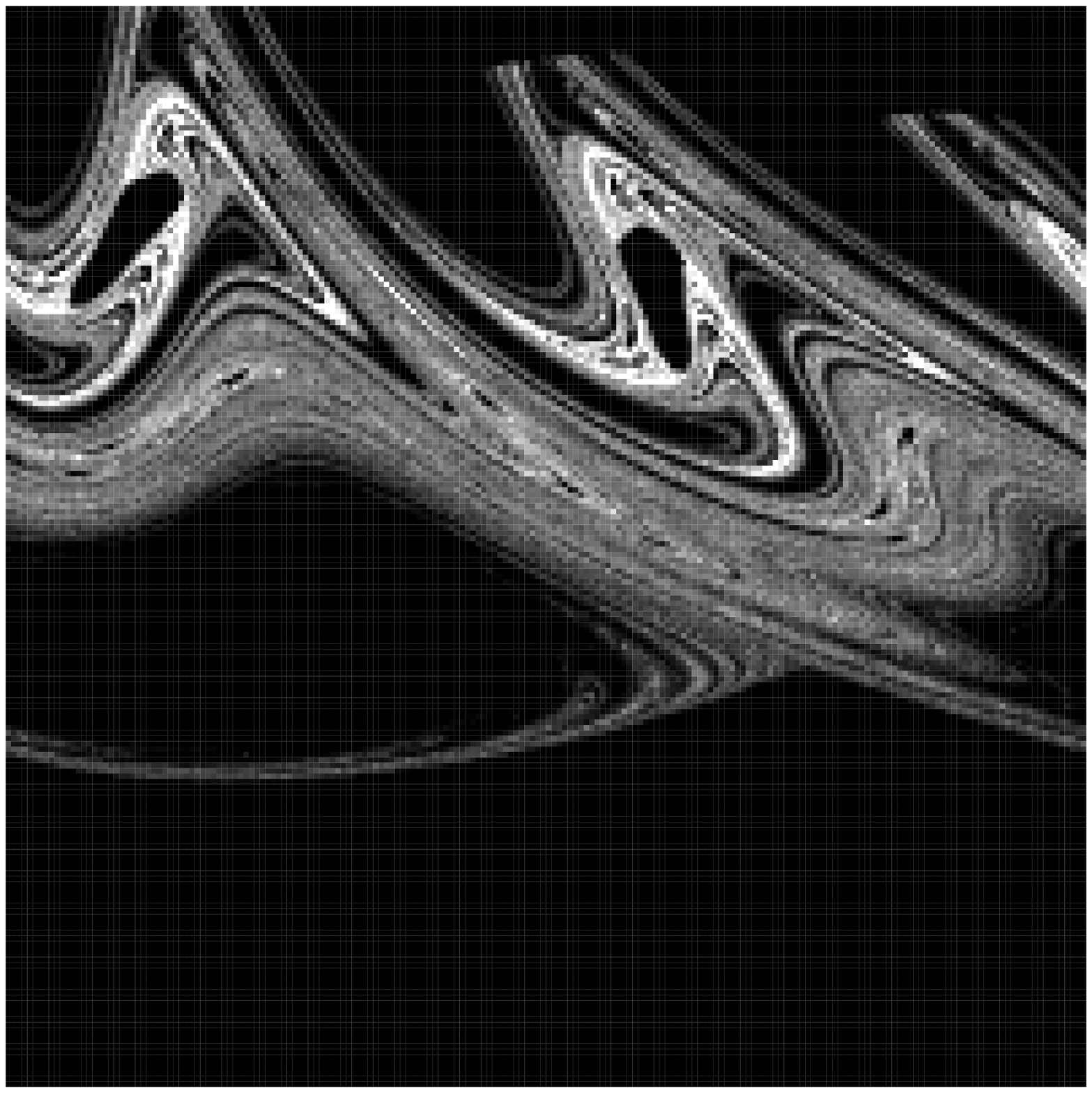}
\epsfxsize=5.0cm\epsffile{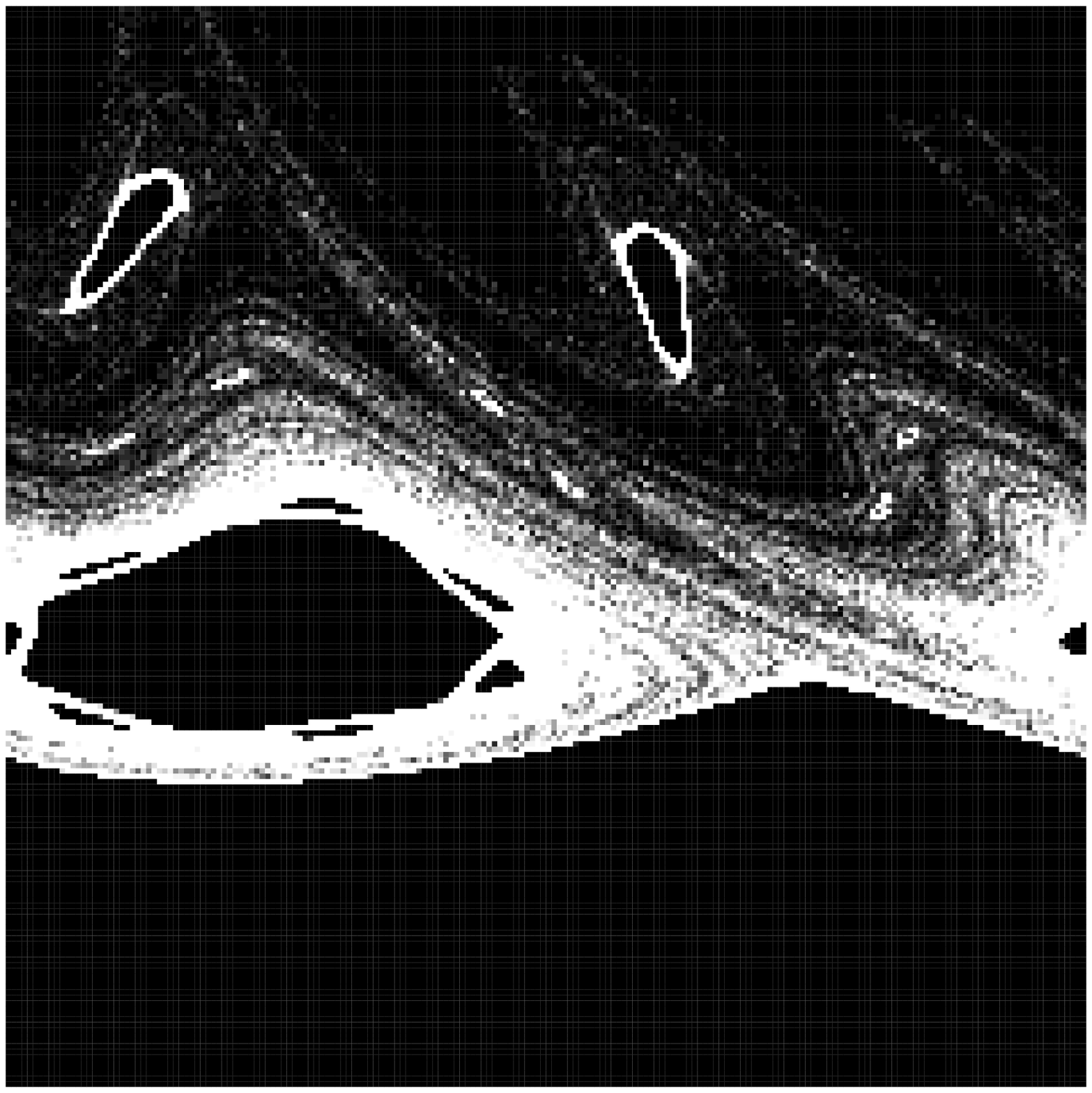}}  
\vspace{-1.9cm}
\centerline{\epsfxsize=5.0cm\epsffile{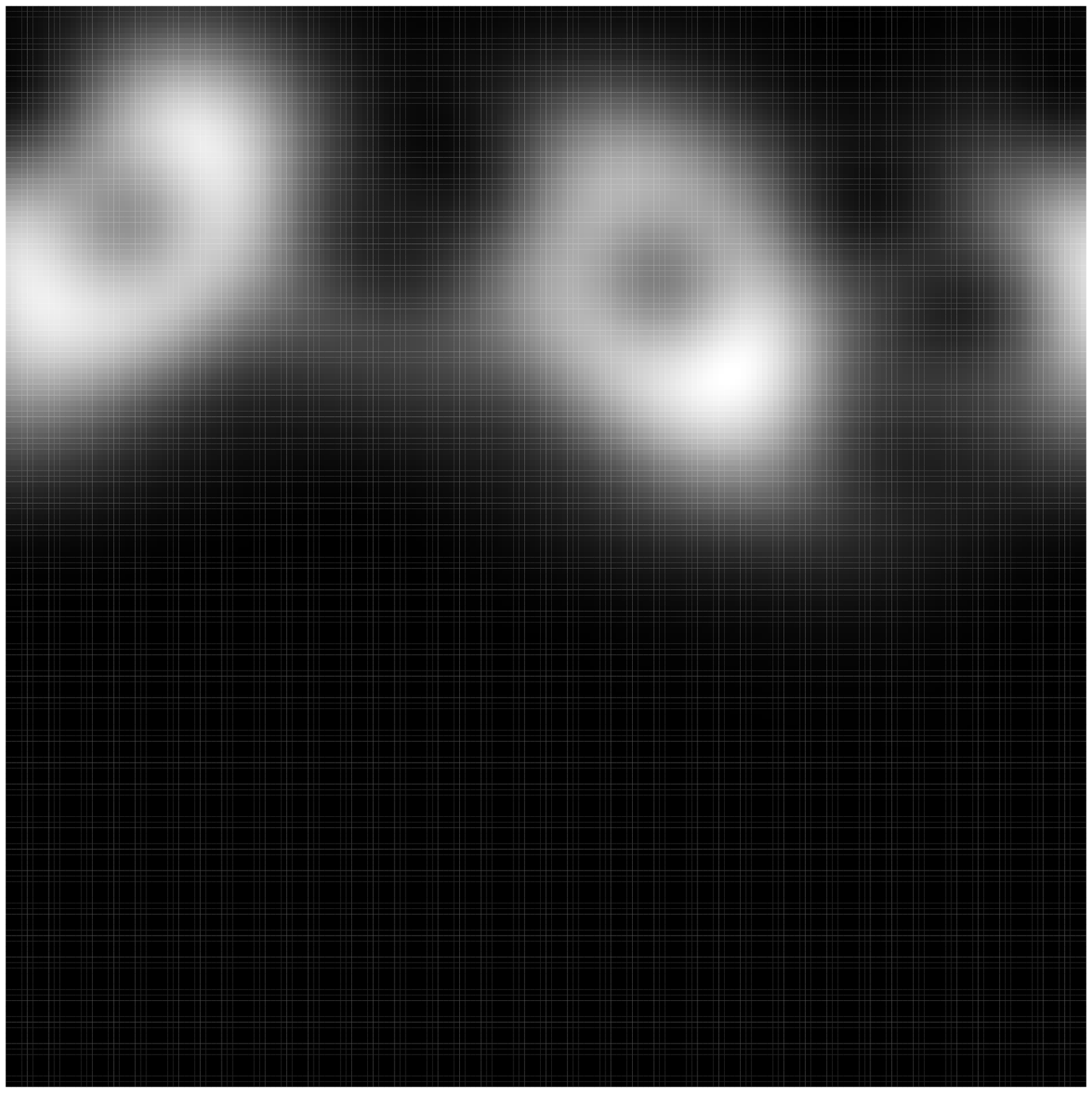}  
\epsfxsize=5.0cm\epsffile{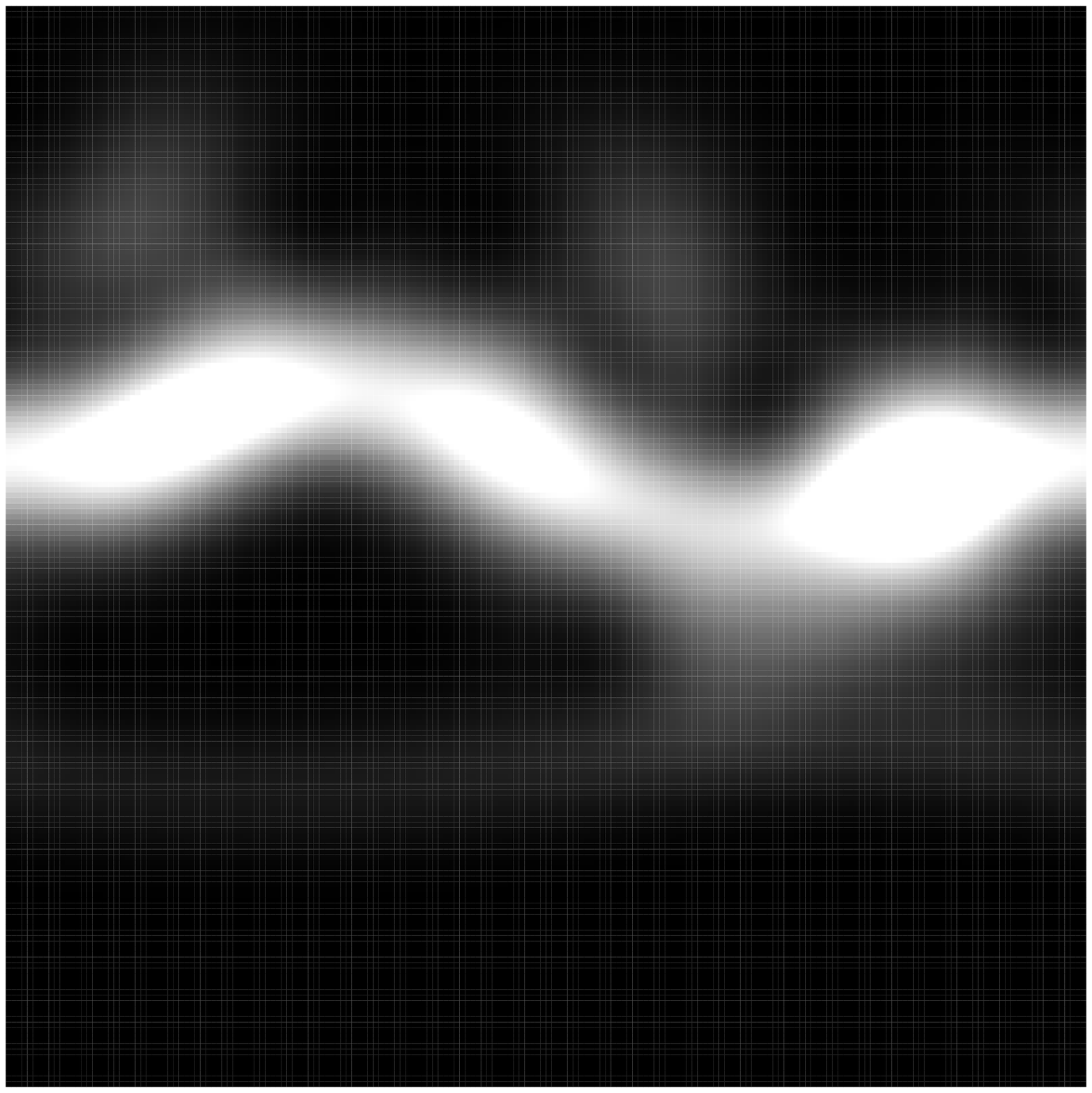}}   
\vspace{-1.9cm}
\centerline{\epsfxsize=5.0cm\epsffile{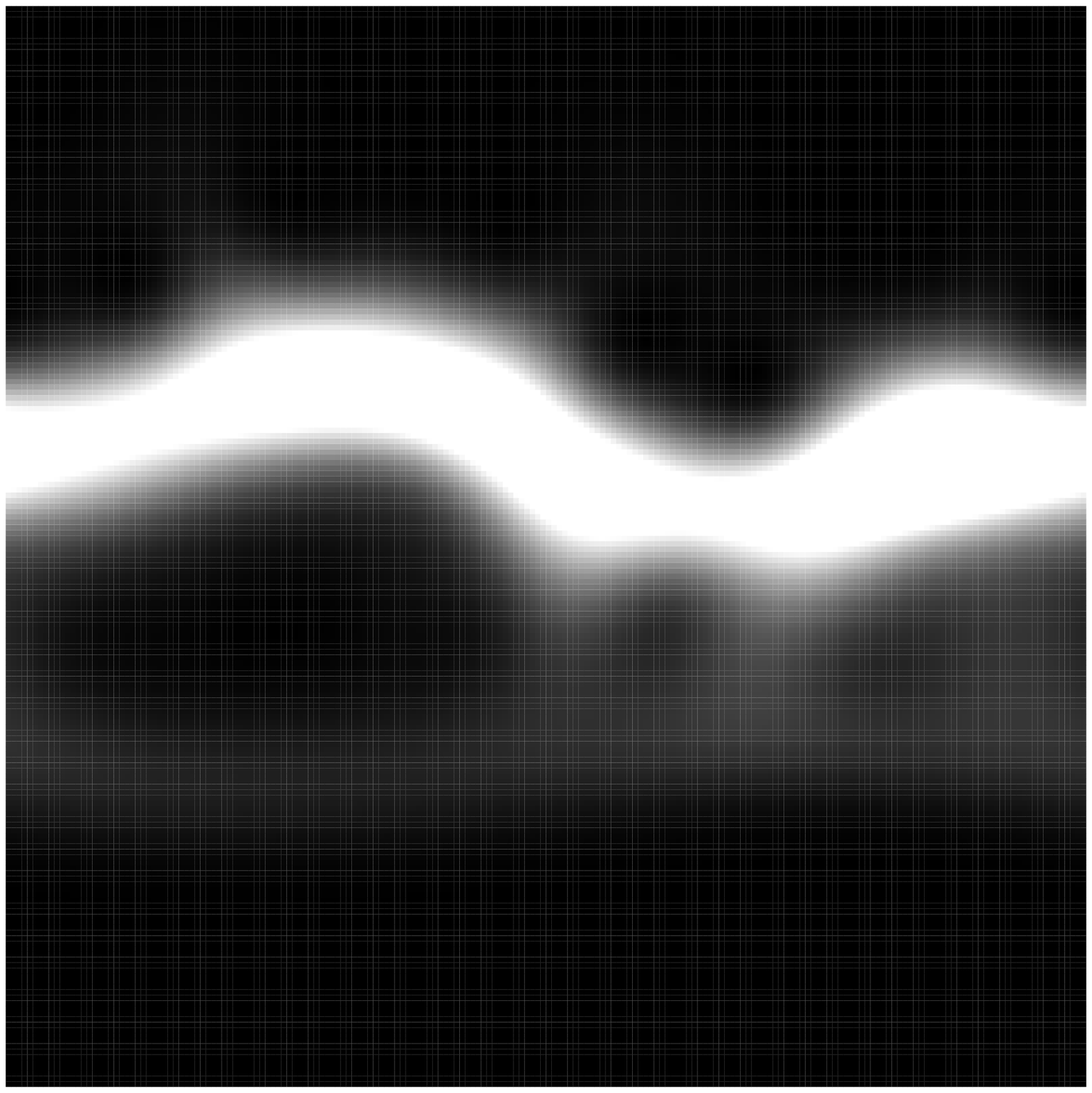}  
\epsfxsize=5.0cm\epsffile{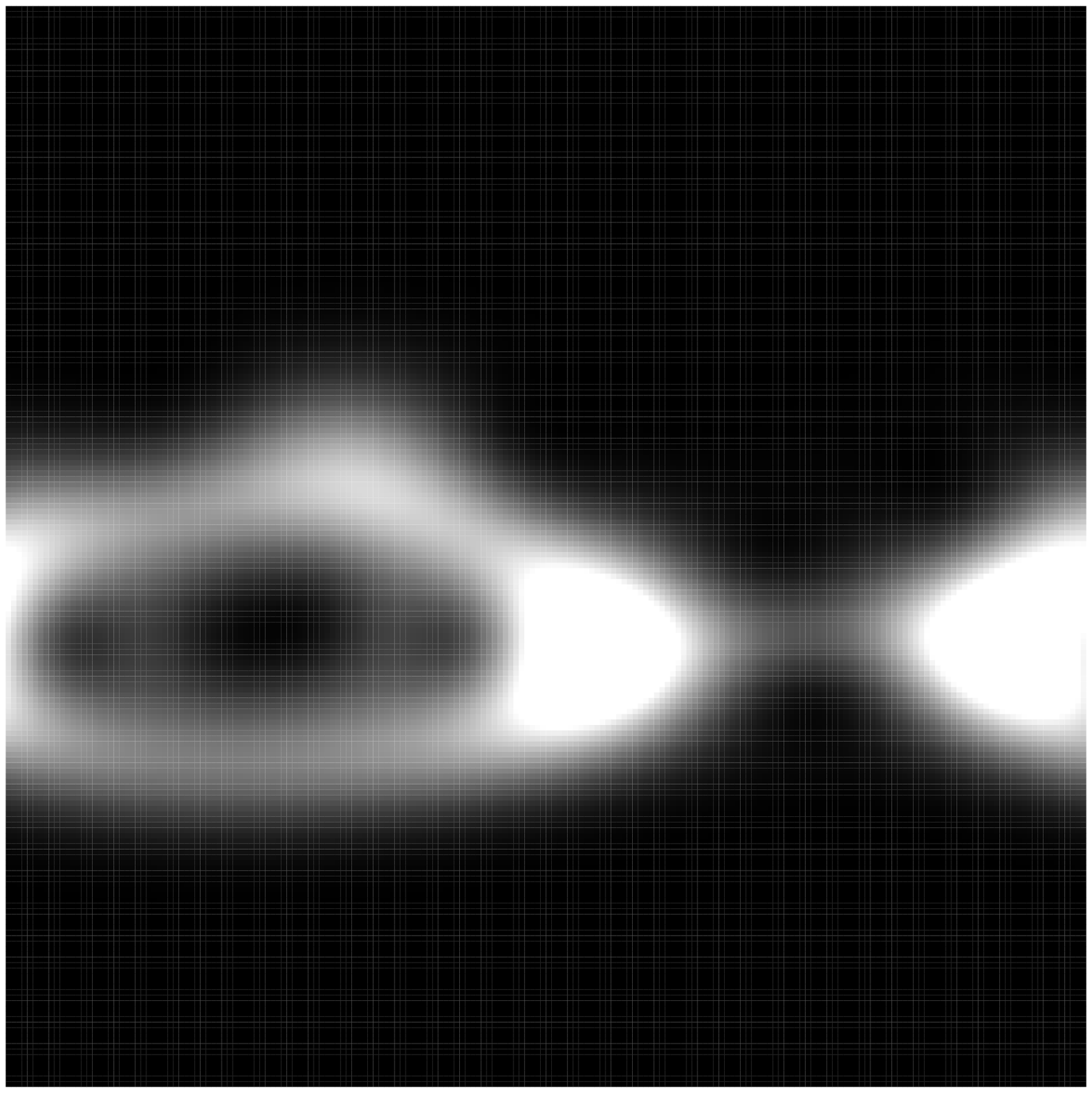}}   
\caption{\small Classical density plot (top) and Husimi function 
(middle and bottom) in action-angle variables ($n,\theta$), with 
$30\leq n\leq 63.5$ (vertical axis) and 
$0\leq\theta<2\pi$ (horizontal axis),   
for the $1d$ model in the case of Fig. \ref{fig1}.  
Top left: $50\leq t \leq 60$; top right:  
$2\times 10^3\leq t \leq 10^4$; 
Husimi functions are averaged in a small time
interval $\delta t=10$ to decrease fluctuations, 
near $t=50$ (middle left), $t=10^3$ (middle right), 
$t=10^4$ (bottom left) and $t=10^5$ (bottom right).  
Bright regions correspond to maximal density.} 
\label{fig2} 
\end{figure} 

In order to confirm that the algebraic probability decay is related 
to localization effects and to the sticking of classical trajectories 
and of quantum 
probability near the integrable islands in the phase space, we show 
in Fig. \ref{fig2} the time evolution of the survival probability 
distribution in the phase space of action-angle variables $(n,\theta)$
for the $1d$ model. 
In the classical case $3\times 10^6$ orbits are initially homogenously 
distributed in the angle $\theta$ on the line $n=n_0=60$, corresponding 
to the initial quantum state with principal quantum number $n_0=60$. 
After $50$ microwave periods, the classical distribution of 
non ionized orbits shows a fractal structure which surrounds 
the stability islands (Fig. \ref{fig2} top left). 
At larger times this distribution approaches 
more and more closely the boundary critical invariant curves
(Fig. \ref{fig2} top right). 
One of them confines the motion in the region with 
$n>n_b\approx n_0 (\epsilon_c/\epsilon_0)^{1/5} \approx 41$
where $n_b$ determines the classical chaos border
for given $\epsilon_0$. 
Other invariant curves mark the critical boundaries around internal 
stability islands (for example at $n\approx 55$, corresponding to 
$\omega n^3\approx 2$).   
This phase space metamorphosis is associated with a change in the 
exponent of the power law decay of classical Poincar\'e recurrences
(the crossover occurs at $t\approx 5\times 10^2$).   
In the quantum case the value of $\hbar_{\rm eff}$ is not sufficiently 
small to resolve the fractal structure at small scales. 
However, the Husimi function (constructed in a way similar to 
Ref. \cite{Jensen}) shows similarities with the classical 
probability distribution at $t=50$ (Fig. \ref{fig2} middle left). 
At longer times, the diffusion towards the boundary at $n_b$ is 
slowed down due to localization effects and penetration of the 
quantum probability inside the classical integrable islands.  
At $t=10^3$ (Fig. \ref{fig2} middle right) the quantum probability 
is concentrated in a layer around $n\approx 49$. 
Due to localization effects, 
the Husimi function does not change significantly for a very long 
interaction time (compare with Fig. \ref{fig2} bottom left at 
$t=10^4$). Eventually the probability starts to penetrate very slowly 
inside the main island at $n\approx n_b$ (see Fig. \ref{fig2} bottom 
right at $t=10^5$). 
Therefore tunneling and localization effects are 
responsible for the slow $1/t$ decay of quantum survival probability 
seen in Fig. \ref{fig1}. 
The chaos border starts to influence the dynamics  
only after a very long interaction time because 
its distance from the starting line $n=n_0$ is much greater than 
the localization length $\ell_\phi$: 
$\rho_b=\ell_\phi/ |N_b| \approx 0.35 < 1$, where 
$N_b=(n_0/2\omega_0)(1-n_0^2/n_b^2)$.

\begin{figure}
\centerline{\epsfxsize=10cm\epsffile{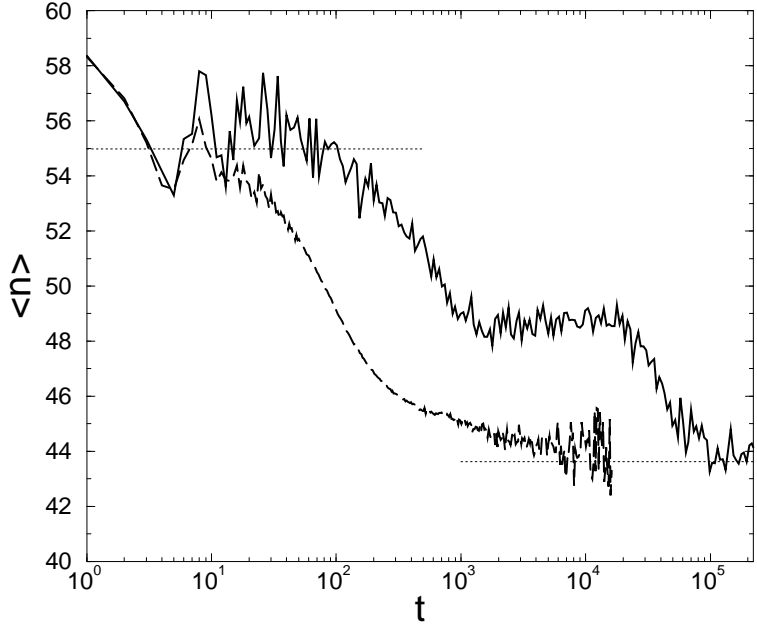}}
\caption{\small Quantum (solid line) and classical (dashed line) 
average action $<n>$ as a function of the interaction 
time $t$ for the case of Fig. \ref{fig1}. The straight lines 
indicate the position of the stability islands for 
$\omega n^3 \approx 1$ ($n\approx 55$) and $\omega n^3\approx 2$ 
($n\approx 44$).}  
\label{fig3} 
\end{figure}

Fig. \ref{fig3} shows the average action $<n>$ of the non-ionized 
electrons as a function 
of the interaction time. The diffusion towards the chaos  
border is slowed down in the quantum case and different regimes 
are clearly visible: tunneling inside the islands for 
$\omega n^3 \approx 2$ (for $t<10^2$), localization of the 
wave packet (for $10^3<t<3\times 10^4$) and tunneling inside 
the main island at $\omega n^3 \approx 1$ (for $t>10^5$). 

Notice that the probability decay $P(t)$ is related to correlations 
decay via $C(t) \propto t P(t) $ \cite{CS99}. 
In the case of $\alpha=1$ this implies that correlations do not 
decay. The Fourier transform of $C(t)$ gives 
the spectral density $S(\omega)$ of the effective noise produced by the 
dynamics: 
\begin{equation}
S(\omega)=\int C(t)\exp(i\omega t)dt\propto 1/\omega. 
\end{equation}
This shows that the spectral noise associated with the quantum 
Poincar\'e recurrences with $\alpha=1$ scales like 
$S(\omega) \propto 1/\omega$. 
A similar behavior of noise has been observed in   
various scientific phenomena \cite{Press}, 
for example in the luminosity 
of stars, the velocity of undersea currents, the flow rate of the 
river Nile, the magnetization of spin glasses and the  
resistance of different solid state devices \cite{noise}.  
It is known as $1/f$ noise and usually extends over  
several orders of magnitude in frequency, indicating a broad distribution of 
time scales in the system. In the case of quantum Poincar\'e 
recurrences this property stems from the exponentially broad 
distribution of escape times from some regions of the phase space, 
due to tunneling and localization effects. 

In summary, we have shown that the survival probability for Rydberg atoms 
in a microwave field decays, up to the Heisenberg time $t_H$, in a way 
similar to the classical probability. 
For $t>t_H$ the quantum probability starts to 
decay slower than the classical one, with the exponent of the 
algebraic decay $\alpha=1$. 
We have given parameter values which should allow one to observe  
quantum Poincar\'e recurrences in microwave experiments with Rydberg atoms. 
Indeed, the thermal beams used with alkali Rydberg atoms allow one 
to vary the interaction time by orders of magnitude up to $10^5$ 
microwave periods \cite{Delande}.   
Therefore the results of the present paper show that theoretical 
and experimental studies of chaotic Rydberg atoms still represent 
a challenge for fundamental research of quantum chaos. 

\vspace{0.5cm}  
\noindent 
{\small Support from the Progetto Avanzato INFM ``Quantum transport 
and classical chaos'' is gratefully acknowledged.}

%
\end{document}